\documentclass[11pt]{article}
\usepackage{amsfonts}
 \usepackage{amssymb}
 \usepackage{amsmath}
 \newfont{\bbbold}{msbm10}

 \def\cF{{\cal F}}

 \def\cO{{\cal O}}

 \newfont{\goth}{eufm10 scaled \magstep1}

 \def\a{\alpha}
 \def\b{\beta}
 \def\c{\gamma}
 \def\d{\delta}\def\D{\Delta}
 \def\e{\epsilon}

 \def\l{\lambda}
 \def\m{\mu}
 \def\n{\nu}

 \def\r{\rho}
 \def\s{\sigma}
 
 \def\th{\theta}
 
 \def\adt{\dot \alpha}
 \def\bdt{\dot \beta}
 \def\cdt{\dot \gamma}
 \def\ddt{\dot \delta}

\def\rdt{\dot \rho}
 \def\sdt{\dot \sigma}
 \def\mdt{\dot \mu}
 \def\ndt{\dot \nu}

 \def\be{\begin{equation}}\def\ee{\end{equation}}
 \def\bea{\begin{eqnarray}}\def\eea{\end{eqnarray}}
 \def\ba{\begin{array}}\def\ea{\end{array}}

 \def\del{\partial}

 \def\str{\rm str}

 \def\del{\partial}

 \def\3dt{\dot{3}}

 {}

 \def\bd{\begin{document}}
 \def\ed{\end{document}}
 \def\bea{\begin{eqnarray}}
 \def\ba{\begin{array}}\def\ea{\end{array}}
 \def\eea{\end{eqnarray}}
 \def\ft#1#2{{\textstyle{{\scriptstyle #1}\over {\scriptstyle #2}}}}
 \def\fft#1#2{{#1 \over #2}}
 \newcommand{\eq}[1]{(\ref{#1})}
 \def\eqs#1#2{(\ref{#1}-\ref{#2})}
 \def\det{{\rm det\,}}
 \def\tr{{\rm tr}}\def\Tr{{\rm Tr}}
  \def\str{{\rm str}} \def\diag{{\rm diag}}
 \def\sdet{{\rm sdet}}\def\symtr{{\rm symtr}}

\newcommand{\hoch}[1]{$^{#1}$}

\def\appendix{{\newpage\section*{Appendices}}\let\appendix\section%
        {\setcounter{section}{0}
        \gdef\thesection{\Alph{section}}}\section} 
\renewcommand{\theequation}{\thesection.\arabic{equation}}
\csname @addtoreset\endcsname{equation}{section}
\begin{document}

\begin{titlepage}
{\hbox to\hsize{${~}$\hfill }}
%{\hbox to\hsize{${~}$ \hfill {CERN--TH/2002-???}}} 
{\hbox to\hsize{${~}$ \hfill {LAPTH-1165/06}}} \vskip 0.2pt 
%\hfill {\tt hep-th/0600000}

\begin{center}
\vglue .3in {\Large\bf Superconformal Invariants or How to Relate
  Four-point AdS Amplitudes}
%\\
%\today

\vskip 1cm

\large{$\rm{J.M. ~Drummond}$}, \large{$\rm{L. ~Gallot}$}
and  \large{$\rm{E.~Sokatchev}$}
\\[.3in]
\small

%\\
%[.03in] $^{\rm(c)}${\it CERN, Theory Division, 1211 Geneva 23, Switzerland}

%[.03in] 
{\it Laboratoire d'Annecy-le-Vieux de Physique 
Th\'{e}orique\footnote[1]{UMR 5108 associ{\'e}e {\`a}
 l'Universit{\'e} de Savoie}
LAPTH, B.P. 110, \\ F-74941 Annecy-le-Vieux et l'Universit\'{e} de Savoie}
\\[.3in]
\normalsize

{\bf ABSTRACT}\\[.0015in]
\end{center}
Using the form of $N=2$ superconformal invariants we derive the explicit
relation between the bottom and top components of the correlator of four
stress-tensor multiplets in $N=4$ Super Yang-Mills. 
The result is given in terms
of an eighth order differential operator acting on the function of
two variables which characterises these correlators. It allows us to show a
non-trivial consistency relation between the known results for the
corresponding supergravity amplitudes on $AdS_5$.

\vskip 7pt ${~~~}$ \newline
%\\[.01in]

\end{titlepage}    

\section{Introduction}
In the context of the AdS/CFT correspondence
\cite{Maldacena:1997re,Witten:1998qj,Gubser:1998bc} 
four-point functions of stress-tensor multiplets in $N=4$ Super
Yang-Mills theory have 
been the subject of many studies, both in perturbation theory
\cite{Eden:1998hh,Eden:1999kh,Bianchi:1999ge,Eden:2000mv,Bianchi:2000hn,Bianchi:2001cm}
and at strong coupling
\cite{Liu:1998ty,Freedman:1998bj,D'Hoker:1999jc,D'Hoker:1999pj,Arutyunov:1999fb,Arutyunov:2000py}.
They exhibit many  
interesting properties in both regimes and contain much information about the
structure of Yang-Mills theory itself. 
%In particular the four-point functions
%of four $N=4$ supercurrents have been the subject of many studies and have
%been directly computed in perturbation theory up to two-loops \cite{} and in
%the supergravity approximation \cite{}. 
These results have been used to
verify the consistency of the operator product expansion
 and to 
calculate the anomalous dimensions of operators of twist two
\cite{Arutyunov:2000ku,Arutyunov:2001mh,Dolan:2001tt,Dolan:2004iy}.

In \cite{Eden:1998hh} the four-point function of four $N=4$ supercurrent
multiplets was studied using $N=2$ harmonic superspace. The $N=4$
multiplet can be
decomposed into $N=2$ multiplets whose calculation can then be performed
using $N=2$ harmonic superspace Feynman rules. 
The first loop correction to the four-point function was evaluated this way
and after various manipulations \cite{Eden:1999kh} the result was
given in terms of a 
single function of the conformally invariant cross-ratios - the well-known
one-loop box integral. It was subsequently shown in general that the full
four-point function of four $N=4$ stress-tensor multiplets is determined by a
single function of the two cross-ratios \cite{Eden:2000bk}. 

In particular, the four-point function of two chiral and two anti-chiral $N=2$
field strength bilinears is sufficient to determine this function completely.
It was shown in \cite{Eden:1998hh} that in principle the full
$N=2$ four-point function is uniquely fixed by superconformal
invariance and the knowledge of the leading term of its theta
expansion. 
It is this part of the $N=4$ four-point function which
contains the correlators of the operators which couple to the dilaton
and axion fields in the supergravity limit.
The aim of this paper is to provide the direct relation between the
top component and the bottom component of this $N=2$ correlation function.  

As is to be expected the top and bottom components are related by a
differential operator which is determined by the structure of the $N=2$
superconformal extension of the two conformal invariant
cross-ratios. We fix the form of this operator in section \ref{calc}.
The final result is remarkably elegant and we present it
here. We consider the $N=2$ correlator of two chiral and two
anti-chiral SYM field strength bilinears with all odd variables set to
zero (a trace over the gauge group for each bilinear is assumed),
\be
\langle W_1^2 \bar{W}_2^2 W_3^2 \bar{W}_4^2 \rangle |_{\th=0} 
= \frac{1}{x_{13}^4 x_{24}^4}F(u,v),
\ee
with cross-ratios
\be
u=\frac{x_{12}^2 x_{34}^2}{x_{13}^2 x_{24}^2}, 
\hspace{20pt}
v=\frac{x_{14}^2 x_{23}^2}{x_{13}^2 x_{24}^2}.
\label{uandv}
\ee
The $\th^4$ component of $W^2$ is $L^+=F_{\a \b}F^{\a \b} + ...$,
where $F_{\a \b}$ is the self-dual part of the Yang-Mills field
strength.
Similarly we write the $\bar{\th}^4$ component of $\bar{W}^2$ as $L^-$.
The top component of the correlator in all odd variables is given by
\be
\langle L^+_1 L^-_2 L^+_3 L^-_4 \rangle  
= \frac{1}{x_{13}^8 x_{24}^8} (\D^{(2)})^2 u^2 v^2
(\D^{(2)})^2 F(u,v) 
\ee
with
\be
\D^{(2)}=u\del_u^2 + v\del_v^2 + (u+v-1)\del_u \del_v + 2(\del_u
+\del_v). \label{opsec}
\ee
The operator (\ref{opsec}) has appeared before in the context of
supersymmetric Ward identities \cite{Eden:2000bk}.

In section \ref{apps} we provide a consistency check on the results of
\cite{D'Hoker:1999pj} and \cite{Arutyunov:2000py} where the AdS/CFT dual
correlators were studied in the supergravity limit.  

\section{N=2 superconformal four point invariants} \label{calc}

We briefly describe various features of the correlator of four $N=4$
stress-tensor multiplets and summarise the known results about its
evaluation in perturbation theory and in the supergravity
approximation. The correlator is
\be
G^{(N=4)}=\langle T^{A_1 B_1} T^{A_2 B_2} T^{A_3 B_3} T^{A_4 B_4} \rangle.
\ee

We follow the notation of \cite{Eden:1998hh} where the operator
$T^{AB}$ is given 
in terms of the $N=4$ field strength superfield $W^A$ by
\be
T^{AB}=W^A W^B -\frac{1}{6} \delta^{AB} W^C W^C,
\ee
with a trace over the gauge group assumed but not written explicitly.

The indices $A,B$ are $SO(6)$ vector indices and there are six different
$SO(6)$ structures in the four-point function. For the bottom component of the
correlator we therefore have
\begin{align}
G^{(N=4)}|_{\th=0} =  \,\, &a_1(u,v)
  \frac{(\delta_{12})^2(\delta_{34})^2}{x_{12}^4 
  x_{34}^4} + a_2(u,v) \frac{(\delta_{13})^2(\delta_{24})^2}{x_{13}^4
  x_{24}^2} + a_3(u,v) \frac{(\delta_{14})^2(\delta_{23})^2}{x_{14}^4
  x_{23}^4} \notag \\ 
+&b_1(u,v) \frac{\delta_{13} \delta_{24} \delta_{14}
\delta_{23}}{x_{13}^2 x_{24}^2 x_{14}^2 x_{23}^2} + b_2(u,v) \frac{\delta_{12}
  \delta_{34} \delta_{14} \delta_{23}}{x_{12}^2 x_{34}^2 x_{14}^2 x_{23}^2}
  +b_3(u,v) \frac{\delta_{12} \delta_{34} \delta_{13} \delta_{24}}{x_{12}^2
  x_{34}^2 x_{13}^2 x_{24}^2}.
\end{align}

Here $u$ and $v$ are the conformally invariant cross-ratios (\ref{uandv})
%\be
%u=\frac{x_{12}^2 x_{34}^2}{x_{13}^2 x_{24}^2}, 
%\hspace{20pt} 
%v=\frac{x_{14}^2 x_{23}^2}{x_{13}^2 x_{24}^2},
%\ee
and 
\be
(\d_{12})^2 (\d_{34})^2 = \d_{\{A_1 B_1 \}}^{A_2 B_2} \d_{\{A_3 B_3\}}^{A_4
  B_4}, \hspace{20 pt} \d_{13}\d_{24}\d_{14}\d_{23} = \d_{\{A_1 B_1\}}^{\{A_3
  \{ B_4 } \d_{\{A_2 B_2\}}^{A_4 \} B_3\}}.
\ee

The functions $a$ and $b$ above are not all independent. Imposing invariance
under the crossing symmetries, which permute the points $(1,2,3,4)$, yields
the following relations,  
\begin{align}
&a_1(u,v)=a_3(v,u)=a_1(u/v,1/v), \notag \\
&a_2(u,v)=a_2(v,u)=a_3(u/v,1/v), \notag \\
&b_1(u,v)=b_3(v,u)=b_1(u/v,1/v), \notag \\
&b_2(u,v)=b_2(v,u)=b_3(u/v,1/v). 
\end{align}
Thus after imposing the crossing symmetry
only one of the $a$ and one of the $b$ are independent. In fact it was shown
in \cite{Eden:2000bk} by considering in detail the quantum corrections to the
correlator that the functions $a$ and $b$ are also related to each other.
The final result of this is that all six parts of the correlator are
determined by their values in free field theory (corresponding to the
constants $A_0$ and $B_0$ below) together with a single
function $\cF$ of the two cross-ratios which encodes the quantum
corrections, 
\begin{align}
a_1(u,v) &= A_0 + u \cF(u,v), \notag \\
a_2(u,v) &= A_0 + \cF(u,v), \notag \\
a_3(u,v) &= A_0 + v \cF(u,v), \notag \\
b_1(u,v) &= B_0 +(u-v-1) \cF(u,v), \notag \\
b_2(u,v) &= B_0 +(1-u-v) \cF(u,v), \notag \\
b_3(u,v) &= B_0 +(v-u-1) \cF(u,v). \label{crossings}
\end{align}
The function $\cF$ satisfies the crossing conditions,
\be
\cF(u,v)=\cF(v,u)=\frac{1}{u}\cF(\frac{1}{u},\frac{v}{u}).
\ee
This result was confirmed in \cite{Dolan:2000ut,Heslop:2002hp} by
imposing consistency of superconformal invariance with the the operator
product expansion. 

In \cite{Eden:1998hh} the splitting of the $N=4$ field strength
multiplet into an 
$N=2$ hypermultiplet and field strength multiplet was described. Explicitly
one can see that the ${\bf 6}$ of $SO(6)$ splits into a ${\bf 3} + {\bf
  \bar{3} }$ of $SU(3)$ which further decompose into a ${\bf 2} + {\bf 1} +
{\bf \bar{2}}+{\bf \bar{1}}$ of $SU(2)$:
\be
W^A \longrightarrow (\phi^i, W , \bar{\phi}_i , \bar{W}).
\ee  
The reference \cite{Eden:1998hh} describes how the $a$ and $b$ functions (or
equivalently $\cF$) can be determined by a set of correlators of $N=2$
hypermultiplet bilinears. It was 
also shown that the possible independent correlators of the $N=2$ field
strength bilinears are all determined by the $a$ and $b$ functions. Explicitly
we have,
\be
\langle W_1^2 W_2^2 W_3^2 W_4^2 \rangle = \langle W_1^2 W_2^2 W_3^2
\bar{W}_4^2 \rangle = 0, \label{triv} 
\ee
while for the bottom component of the only non-trivial correlator of this type
we have,
\be
\langle W_1^2 \bar{W}_2^2 W_3^2 \bar{W}_4^2 \rangle |_{\th=0} =
\frac{a_1(u,v)}{x_{12}^4 x_{34}^4} + \frac{a_3(u,v)}{x_{14}^4
  x_{23}^4} + \frac{b_2(u,v)}{x_{12}^2 x_{34}^2 
  x_{14}^2 x_{23}^2} = \frac{1}{x_{13}^4 x_{24}^4} F(u,v). \label{corr} 
\ee

The function $F$ is symmetric $F(v,u)=F(u,v)$ and is given by
\begin{align}
F(u,v) &= \frac{1}{u^2}a_1(u,v) + \frac{1}{v^2} a_3(u,v) + \frac{1}{uv}
b_2(u,v) \notag \\
&= A_0 \bigl( \frac{1}{u^2} + \frac{1}{v^2} \bigr)+B_0\frac{1}{uv} +
\frac{1}{uv}\cF(u,v).  
\end{align}

%We will study perturbative loop corrections and the connected part of
%the leading $1/N^2$ correction in the supergravity limit. Each of
%these contributes only to the two variable function $\cF(u,v)$. Thus,
%ignoring the free-field  $A_0$ and $B_0$, this becomes
%\be
%F(u,v) = \frac{1}{uv}\cF(u,v). \label{ftocf}
%\ee

The analysis of \cite{Eden:2000bk} shows that $A_0$ and $B_0$ receive
contributions only from free field theory i.e. quantum corrections
contribute only to $\cF$.
The function $\cF$ can be extracted from direct calculations at order
$g^2$ \cite{Eden:1998hh} and order $g^4$
\cite{Eden:2000mv,Bianchi:2000hn} in perturbation theory and at order
$1/N^2$ in the supergravity approximation \cite{Arutyunov:2000py}.

The full $\th$ expansion of the correlator of two $W^2$ and two
$\bar{W}^2$ is in fact determined by the
constraints of $N=2$ superconformal symmetry, as demonstrated in
\cite{Eden:1998hh}. 
%The argument involves extending the conformal invariants 
%to superconformal invariants and then arguing
%that there can be no nilpotent invariants (as long as the four
%positions $(x_1,x_2,x_3,x_4)$ are kept separated). 
The top component of the $N=2$ correlator is
\be
\langle
W_1^2 \bar{W}_2^2 W_3^2 \bar{W}_4^2 \rangle |_{\th^{16}} = \langle
L^+_1 L^-_2 L^+_3 L^-_4 \rangle, \label{top}
\ee
where $L^+ = F_{\a \b}F^{\a \b}+...$ and $L^-=\bar{F}_{\adt \bdt}
\bar{F}^{\adt \bdt}+...$ contain the squares of the self-dual and 
anti-self-dual parts of the Yang-Mills field strength,
\be
F_{\m \n} = (\s_{\m \n})^{\a \b}F_{\a \b} + (\tilde{\s}_{\m \n})^{\adt
  \bdt}\bar{F}_{\adt \bdt}.
\ee
{}From the knowledge of the correlators (\ref{triv}, \ref{top}) one can 
construct any four-point correlator of the operators $F_{\m \n}F^{\m
  \n}+...$ and  $F_{\m \n} \tilde{F}^{\m \n}+...$ which correspond to 
the four-point supergravity amplitudes of dilaton and axion fields in
the supergravity limit.

We now briefly review the construction of superconformal invariants
\cite{Eden:1998hh,Heslop:2002hp}. 
The $N=2$ superspace has
$(x^{\alpha\dot\alpha},\theta^{\alpha}_i,\bar\theta^{\dot\alpha i} )$ as
coordinates.  
Since the correlation function we are interested in is chiral at points 1 and
3 and anti-chiral at points 2 and 
4, it depends only on the chiral or anti-chiral coordinates at these
points. These are given by 
\begin{equation}
  \begin{array}{lclcll}
    X^{\alpha\dot\alpha}&=&x^{\alpha\dot\alpha}+
    2{i}\theta^{\alpha}_i\bar\theta^{\dot\alpha i}
    &;&\theta^{\alpha}_i &\text{chiral} \\
    \bar X^{\alpha\dot\alpha}&=&x^{\alpha\dot\alpha}-2{i}\theta^{\alpha}_i
    \bar\theta^{\dot\alpha i}
    &;&\bar\theta^{\dot\alpha i}&\text{anti-chiral}
  \end{array}
\end{equation}

One then forms Q-supersymmetric chiral-anti-chiral differences
\be
y_{r \bar s}=X_r-\bar{X}_{\bar s} - 4i \th_r \cdot \bar{\th}_{\bar s}.
\ee
S-supersymmetry induces a $GL(2)\times GL(2)$ transformation on these
quantities, 
\be
\d y_{r \bar s}^{\a \adt} = 
\th_{ri}^{\a} \eta_{\b}^{i} y_{r \bar s}^{\b \adt} 
+ y_{r \bar s}^{\a \bdt}\bar{\eta}_{i \bdt}\bar{\th}_{\bar s}^{i \adt}.
\ee
To obtain invariants we should then form $GL(2)\times GL(2)$ invariant
combinations of the $y_{r \bar s}$. Following logic similar to that of
\cite{Heslop:2002hp} we find that the two independent such quantities can be
written as $\tr Z$ and $\det Z$ where $Z$ is the $2\times 2$ matrix
given by
\be
Z^{\adt}{}_{\bdt} = 
(y_{1 \bar 2}^{\phantom{-1}} y_{1 \bar 4}^{-1} 
y_{3 \bar 4}^{\phantom{-1}} y_{3 \bar 2}^{-1})^{\adt}{}_{\bdt}.
\ee
The variables $\hat S$ and $\hat T$ of \cite{Eden:1998hh} are related
to $\tr Z$ and $\det Z$ via
\be
\hat S = \det Z, \hspace{20pt} \hat T = 1 + \det Z - \tr Z = \det(1-Z).
\ee
Equivalently we can work with the superconformal extensions of the
cross ratios $u$ and $v$ given by
\be
\hat U = \frac{\hat S}{\hat T} = \frac{\det Z}{1 + \det Z -\tr Z},
\hspace{20pt} 
\hat V = \frac{1}{\hat T} = \frac{1}{1+ \det Z -\tr Z}.
\ee

Furthermore, as discussed in \cite{Eden:1998hh}, there are no nilpotent
superconformal invariants (as can be seen by counting the number of
odd symmetries and odd variables). 
Thus if the correlator with all variables set to zero is given by
\be
\langle W_1^2 \bar{W}_2^2 W_3^2 \bar{W}_4^2 \rangle |_{\th=0} 
= \frac{1}{x_{13}^4 x_{24}^4}F(u,v)
= \frac{u v F(u,v)}{x_{12}^2 x_{34}^2 x_{14}^2 x_{23}^2}
\ee
then the full correlator is given by its unique superconformal extension,
\be
\langle W_1^2 \bar{W}_2^2 W_3^2 \bar{W}_4^2 \rangle 
= \frac{\hat{U} \hat{V} F(\hat{U},\hat{V})}
{y_{1 \bar 2}^2 y_{3 \bar 4}^2 y_{1 \bar 4}^2 y_{3 \bar 2}^2}.
\ee

All that remains to obtain the $\th^{16}$ component is to expand out
in the odd variables and keep only the top term. 
We keep track of all the contributions using MATHEMATICA. The analysis
is helped by realising that the terms of the form $\th^{16}$ can be
grouped into three types,
\bea
&(\th_r \cdot \bar{\th}_{\bar s})^4 (\th_r \cdot \bar{\th}_{\bar u})^0 
(\th_t \cdot \bar{\th}_{\bar s})^0(\th_t \cdot \bar{\th}_{\bar u})^4 
\hspace{20pt}& {\rm (Type 1)} \notag
\\\notag\\
&(\th_r \cdot \bar{\th}_{\bar s})^3 (\th_r \cdot \bar{\th}_{\bar u})^1 
(\th_t \cdot \bar{\th}_{\bar s})^1(\th_t \cdot \bar{\th}_{\bar u})^3 
\hspace{20pt}& {\rm (Type 2)} \notag
\\\notag\\
&(\th_r \cdot \bar{\th}_{\bar s})^2 (\th_r \cdot \bar{\th}_{\bar u})^2 
(\th_t \cdot \bar{\th}_{\bar s})^2(\th_t \cdot \bar{\th}_{\bar u})^2 
\hspace{20pt}& {\rm (Type 3)} .
\eea
Each of these can be replaced by a tensor constructed from $\e^{\a
  \b}$ and $\e^{\adt \bdt}$ (we give these in the
appendix). Equivalently one can think of applying an odd derivative
with respect to each of the odd variables (sixteen in all). The four
$\th_1$ and four $\th_3$ derivatives induce an eighth order
differential operator on the function $F$. Each $\th$ comes with a
$\bar{\th}$ so the remaining $\bar \th$ derivatives just soak up these
factors and do not increase the order of the operator further.

The final expression we obtain for the operator is remarkably
simple. We find
\be
\langle L^+_1 L^-_2 L^+_3 L^-_4 \rangle  
= \frac{1}{x_{13}^8 x_{24}^8} \D^{(8)} F(u,v) 
\equiv \frac{1}{x_{13}^8 x_{24}^8} H(u,v) 
\ee
with the eighth-order operator given by
\be
\D^{(8)} = (\D^{(2)})^2 u^2 v^2 (\D^{(2)})^2,
\label{operator}
\ee
where
\be
\D^{(2)} = u\del_u^2 + v\del_v^2 + (u+v-1)\del_u
\del_v +2 \del_u + 2\del_v.
\label{op2}
\ee

We can further simplify by employing the variables\footnote{These
  variables or similar ones have appeared in many papers
  \cite{Davydychev:1992mt,Eden:2000bk,Dolan:2000ut,Heslop:2002hp,Dolan:2004mu}.
  They are complex and conjugate to each other in Euclidean space
  while in Minkowski space they are real and independent.} $x$ and
  $\bar x$ where     

\be
u = x \bar x, \hspace{20pt} v=(1 - x)(1 - \bar x).
\ee

In these variables the expression for the second order operator is

\be
\D^{(2)}=(x-\bar x)^{-1}\del_x \del_{\bar x} (x - \bar x)
\ee
and hence the full eighth order operator is
\be
\D^{(8)} = (x-\bar x)^{-1} 
\big[ \del_x^2 x^2(1-x)^2 \del_x^2 \big]
\big[ \del_{\bar x}^2 \bar{x}^2 (1-\bar x)^2 \del_{\bar x}^2 \big]
(x-\bar x).
\ee

Alternatively we can express the operator through the action of
derivatives at each spacetime point,

\be
\frac{1}{x_{13}^8 x_{24}^8} \D^{(8)}F(u,v)= 
\Bigl( 
(x_{13}^2 x_{24}^2)^{-1} \Box_4
\frac{1}{v} \Box_3 v^2 \Box_2 \frac{1}{v} \Box_1 x_{13}^2 x_{24}^2
\Bigr)
\frac{1}{x_{13}^4 x_{24}^4} F(u,v).
\ee

\section{Supergravity Limit} \label{apps}

We now discuss the relation of the results of
\cite{Freedman:1998bj,D'Hoker:1999jc,D'Hoker:1999pj} and
\cite{Arutyunov:1999fb,Arutyunov:2000py} on the supergravity limit of
the correlator under consideration. By the AdS/CFT correspondence
correlators of 
gauge-invariant local operators in the $N=4$ $SU(N)$ Super-Yang-Mills theory
in the limit of large $N$ and strong coupling are given by supergravity
amplitudes. In particular the correlators of the operators $F_{\m \n} F^{\m
  \n}+..$ and $F_{\m \n}\tilde{F}^{\m \n}+...$ are related to dilaton/axion
amplitudes in supergravity according to the identification 
$F_{\m \n}F^{\m \n}+... \sim \cO_\phi$ and 
$F_{\m \n}\tilde{F}^{\m \n}+... \sim \cO_C$ a relation first stressed
in \cite{Liu:1998ty}.  

%The operator we have derived is dependent only on the superconformal
%properties of the correlator (\ref{corr}). Therefore it relates the
%bottom and top component correlators non-perturbatively and so can be
%used equally well in the supergravity limit as in perturbation theory. 
%We write the top component of the correlator of two $W^2$ and two
%$\bar{W}^2$ as
%\be
%\langle M^2_1 \bar{M}^2_2 M^2_3 \bar{M}^2_4 \rangle =
%\frac{1}{x_{13}^8 x_{24}^8} H(u,v),
%\ee
%where we consider contributions to the function $H$ given by 
%\be
%H(u,v) = (\D^{(2)}_{uv})^2 u^2 v^2 (\D^{(2)}_{uv})^2 \frac{1}{uv} \cF(u,v).
%\label{Hfun}
%\ee
%Here $\D^{(2)}_{uv} = u\del_u^2 + v\del_v^2 + (u+v-1)\del_u \del_v
%+2(\del_u+\del_v)$. 

To obtain the correlators of scalar composites built from the
Yang-Mills field strength and its dual we use the relations
\begin{align}
L^+ + L^- = F_{\m \n} F^{\m \n} + ... 
\notag \\
iL^+-iL^- = F_{\m \n} \tilde{F}^{\m \n} +...
\end{align}

Recalling (\ref{triv}) we find that
\begin{align}
&\langle 
(F_{\m \n} F^{\m \n}+...)_1 (F_{\m \n} \tilde{F}^{\m \n}+...)_2 
(F_{\m \n} F^{\m \n}+...)_3 (F_{\m \n} \tilde{F}^{\m \n}+...)_4 
\rangle \notag \\
&\notag \\
=&\langle L^+_1 L^-_2 L^+_3 L^-_4 \rangle
-(2\leftrightarrow 3) - (1\leftrightarrow 2)
\label{ffffcorr}
%-\langle M^2_1 \bar{M}^2_3 M^2_2 \bar{M}^2_4 \rangle
%-\langle M^2_2 \bar{M}^2_1 M^2_3 \bar{M}^2_4 \rangle .
\end{align}

In terms of $H(u,v) \equiv \D^{(8)}F(u,v)$ expression (\ref{ffffcorr}) becomes

\be
\frac{1}{x_{13}^8 x_{24}^8} 
\Bigl( 
H(u,v) 
-\frac{1}{u^4} H\bigl( \frac{1}{u},\frac{v}{u} \bigr)
-\frac{1}{v^4} H\bigl( \frac{u}{v},\frac{1}{v} \bigr)
\Bigr). \label{Hcombos}
\ee

Thus, given an expression for the function $F$ we can determine the
function $H$ through the action of the differential operator and hence
the correlator (\ref{ffffcorr}).

In \cite{Freedman:1998bj,D'Hoker:1999pj} the calculation of dilaton/axion
amplitudes in IIB supergravity on $AdS_5\times S^5$ was performed. The
final result is expressed in terms of the so-called D-functions. These
are functions of the six squares of differences of the 
four points $x_1,x_2,x_3,x_4$. The D-functions also appear in the
result of \cite{Arutyunov:2000py} for the bottom component of the same
amplitudes. These functions can all be generated from the one-loop
scalar box function $\Phi^{(1)}(u,v)$ by acting with derivatives with
respect to the square differences. On the other hand
\cite{Eden:2000bk} one can use the fact that the derivatives of
$\Phi^{(1)}$ can be derived from the explicit expression
\cite{Usyukina:1992jd}  
\be
\Phi^{(1)}(u,v)=\frac{1}{\l} \left( 2\bigl( {\rm Li}_2(-\r u) +{\rm Li}_2(-\r
  v) \bigr)+\ln \frac{v}{u} \ln \frac{1+\r v}{1+\r u} + \ln \r u \ln \r v +
  \frac{\pi^2}{3} \right), 
\ee
where
\be
\l(u,v)=\sqrt{(1-u-v)^2 -4uv}, \hspace{20pt} \r(u,v)=2(1-u-v+\l)^{-1}. 
\ee
This implies that the derivatives of $\Phi^{(1)}$ can be expressed in terms of
$\Phi^{(1)}$ itself and logarithms,
\be
\del_u \Phi^{(1)}(u,v) = \frac{1}{\l^2}\left( \Phi^{(1)}(u,v)(1-u+v) +
  2\ln u - (u+v-1)\frac{\ln v}{u} \right) \label{delphi}
\ee
and similarly for $u\longleftrightarrow v$.

Thus any expression given in terms of D-functions can be translated into an
expression involving $\Phi^{(1)}$ and logarithms. The expression for
the function defining the bottom component of the amplitude from
\cite{Arutyunov:2000py} corresponds to (see \cite{Eden:2000bk})

\be
\frac{a_1(u,v)}{x_{12}^4 x_{34}^4} 
= \frac{32 N_c^2}{2^8 \pi^{10}}
\Bigl(  -\frac{1}{2} \frac{D_{2211}}{x_{34}^2} + \bigl( \frac{1}{u} +
    \frac{v}{u} -1 \bigr) x_{12}^2 D_{3322} + \frac{3}{2} D_{2222}
    \Bigr). 
\label{afa1}
\ee

which can be written as a third order operator acting on
$\Phi^{(1)}(u,v)$. Recall the function $\cF$ is related to quantum
correction part of $a_1$ via $\cF(u,v) = (1/u)a_1(u,v)$. Following
this logic the expression for $\cF(u,v)$ obtained in supergravity is
\cite{Eden:2000bk} 

\begin{align}
\cF_{SG}(u,v) = -\frac{16 N_c^2}{(2 \pi)^8 \l^6} 
\Bigl\{&\Phi^{(1)}(u,v) 12 uv [(1+u+v)\l^2 + 10uv] \notag \\
&+(\ln u) 2u[(1 + v^2 - u - uv +10v)\l^2 + 30 uv(1+v-u)] \notag \\
&+(\ln v) 2v[(1 + u^2 - v - uv +10u)\l^2 + 30 uv(1+u-v)] \notag \\
&+[(1+u+v)\l^4 + 20 uv \l^2] \Bigr\}.
\end{align}

In fact the neatest expression for the third order operator is
\cite{Arutyunov:2000ku} 

\be
\cF_{SG}(u,v) = -\frac{16 N_c^2}{(2 \pi)^8} 
(1 + u\del_u + v\del_v)(uv\del_u \del_v)\Phi^{(1)}(u,v). 
\label{afresultcf}
\ee

{}From this we obtain an expression for the function $H$ as an eleventh
order differential operator acting on $\Phi^{(1)}$.
 
\be
H_{SG}(u,v) = \D^{(8)} F_{SG}(u,v) 
= (\D^{(2)})^2 u^2 v^2 (\D^{(2)})^2 \frac{1}{uv} \cF_{SG}(u,v). 
\ee

This can be rewritten in terms of $\Phi^{(1)}$ itself and
logarithms following the same logic as above. Finally we can form the
combination in equation (\ref{Hcombos}) and use the crossing
properties of $\Phi^{(1)}$ to obtain an expression for the correlator
(\ref{ffffcorr}) in terms of $\Phi^{(1)}$ and logarithms.

The result of \cite{D'Hoker:1999pj} for the connected order $N_c^2$ part of
the dilaton/axion amplitude is 
\begin{align}
\langle \cO_{\phi}(x_1) \cO_C (x_2) \cO_{\phi}(x_3) \cO_C (x_4) \rangle 
=\Bigl(\frac{6}{\pi^2}\Bigr)^4 \Bigl( & 16 \bigl( \frac{s+1}{t} + 3
\bigr) x_{24}^2 D_{4545} + \frac{64}{9} 2 
\frac{s+1}{t} \frac{x_{24}^2}{x_{13}^2} D_{3535} \notag \\
&+ \frac{16}{3}2\frac{s+1}{t} \frac{x_{24}^2}{x_{13}^4} D_{2525} - 14
D_{4444} - \frac{46}{9} \frac{1}{x_{13}^2} D_{3434} \notag \\
&- \frac{40}{9}\frac{1}{x_{13}^4} D_{2424} - \frac{8}{3}
\frac{1}{x_{13}^6} D_{1414} \Bigr). \label{dilatonaxion}
\end{align}

We can follow the same procedure as was used on the expression above
for the bottom component to express this in terms of $\Phi^{(1)}$ and
logarithms. Comparing this to the expression obtained from the action
of our eighth order operator on the result of \cite{Arutyunov:2000ku}
we find perfect agreement up to an overall constant factor dependent
on conventions,

\begin{align}
\langle \cO_{\phi}(x_1) \cO_C (x_2) \cO_{\phi}(x_3) \cO_C (x_4) \rangle 
=\frac{1}{x_{13}^8 x_{24}^8} 
\Bigl( 
H_{SG}(u,v) 
-\frac{1}{u^4} H_{SG}\bigl( \frac{1}{u},\frac{v}{u} \bigr)
-\frac{1}{v^4} H_{SG}\bigl( \frac{u}{v},\frac{1}{v} \bigr)
\Bigr). 
\end{align}

This is a highly non-trivial check of the compatibility
of the results of \cite{D'Hoker:1999pj} and \cite{Arutyunov:2000py}
with superconformal symmetry.  

\section{Summary}
We have given here the explicit relation between the bottom and top
components of the correlator of two chiral and two anti-chiral $N=2$
field strength bilinears. The relation follows from $N=2$
superconformal invariance as discussed in \cite{Eden:1998hh}. The
result is expressed in terms of an eighth order differential operator
acting on a function of two variables (the conformal
cross-ratios). The same method can be used to obtain differential
operators for correlators of higher weights.
The operator we have derived has allowed us to establish the equivalence of
known results \cite{D'Hoker:1999pj,Arutyunov:2000py} for supergravity
amplitudes.            
The methods we have described here could prove useful in studying the
supergravity amplitudes beyond the leading $1/N^2$ correction. 

{\bf Acknowledgements} E. S. acknowledges Hugh Osborn for a discussion
which stimulated this research.

\appendix{$\theta$-identities}\label{AppA}

We give here the identities for the product of eight indexed $\theta$s
and eight indexed $\bar\theta$s in terms of $\Theta^{16}$ up to some
tensorial structure. The tensors $\e^{\a \b}$ and $\e^{\adt \bdt}$ are
antisymmetric with $\e^{12}=\e^{\dot{1} \dot{2}}=1$. 

\begin{align}
&(\th_{r}\cdot \bar{\th}_{\bar s})^{\a \adt}
(\th_{r}\cdot \bar{\th}_{\bar s})^{\b \bdt}
(\th_{r}\cdot \bar{\th}_{\bar s})^{\c \cdt}     
(\th_{r}\cdot \bar{\th}_{\bar s})^{\d \ddt}
(\th_{t}\cdot \bar{\th}_{\bar u})^{\r \rdt}
(\th_{t}\cdot \bar{\th}_{\bar u})^{\s \sdt}
(\th_{t}\cdot \bar{\th}_{\bar u})^{\m \mdt}    
(\th_{t}\cdot \bar{\th}_{\bar u})^{\n \ndt}  \notag \\
&=\Theta^{16} T_1^{\a \adt \b \bdt \c \cdt \d \ddt \r \rdt \s \sdt \m
  \mdt \n \ndt}
\label{c1}
\end{align}
\begin{align}
&(\th_{r}\cdot \bar{\th}_{\bar s})^{\a \adt}
(\th_{r}\cdot \bar{\th}_{\bar s})^{\b \bdt}
(\th_{r}\cdot \bar{\th}_{\bar s})^{\c \cdt}     
(\th_{r}\cdot \bar{\th}_{\bar u})^{\d \ddt}
(\th_{t}\cdot \bar{\th}_{\bar s})^{\r \rdt}
(\th_{t}\cdot \bar{\th}_{\bar u})^{\s \sdt}
(\th_{t}\cdot \bar{\th}_{\bar u})^{\m \mdt}    
(\th_{t}\cdot \bar{\th}_{\bar u})^{\n \ndt}  \notag \\
&=\Theta^{16} T_2^{\a \adt \b \bdt \c \cdt \d \ddt \r \rdt \s \sdt \m
  \mdt \n \ndt}
\label{c2}
\end{align}
\begin{align}
&(\th_{r}\cdot \bar{\th}_{\bar s})^{\a \adt}
(\th_{r}\cdot \bar{\th}_{\bar s})^{\b \bdt}
(\th_{r}\cdot \bar{\th}_{\bar u})^{\c \cdt}     
(\th_{r}\cdot \bar{\th}_{\bar u})^{\d \ddt}
(\th_{t}\cdot \bar{\th}_{\bar s})^{\r \rdt}
(\th_{t}\cdot \bar{\th}_{\bar s})^{\s \sdt}
(\th_{t}\cdot \bar{\th}_{\bar u})^{\m \mdt}    
(\th_{t}\cdot \bar{\th}_{\bar u})^{\n \ndt}  \notag \\
&=\Theta^{16} T_3^{\a \adt \b \bdt \c \cdt \d \ddt \r \rdt \s \sdt \m
  \mdt \n \ndt}
\label{c3}
\end{align}

The three tensors appearing in the above expressions are
\begin{align}
&T_1^{\a \adt \b \bdt \c \cdt \d \ddt \r \rdt \s \sdt \m
  \mdt \n \ndt} = \notag \\
&4(\e^{\a \b} \e^{\adt \bdt} \e^{\c \d} \e^{\cdt \ddt} +  
{\rm cyc}(\b \bdt, \c \cdt,\d \ddt))
(\e^{\r \s} \e^{\rdt \sdt} \e^{\m \n} \e^{\mdt \ndt} +  
{\rm cyc}(\s \sdt, \m \mdt, \n \ndt))
\end{align}
\begin{align}
&T_2^{\a \adt \b \bdt \c \cdt \d \ddt \r \rdt \s \sdt \m
  \mdt \n \ndt}= \notag \\
&2\bigl(
(\e^{\a \b} \e^{\adt \bdt} \e^{\cdt \ddt} \e^{\c \n} 
\e^{\r \s} \e^{\rdt \sdt} \e^{\mdt \ndt} \e^{\d \m} + {\rm cyc}(\s
  \sdt, \m \mdt, \r \rdt)) + {\rm cyc}(\b  \bdt, \c \cdt, \a \adt)  
\bigr)
\end{align}
\begin{align}
&T_3^{\a \adt \b \bdt \c \cdt \d \ddt \r \rdt \s \sdt \m
  \mdt \n \ndt}= \notag \\
&2\bigl(
\e^{\a \b} \e^{\adt \bdt} \e^{\c \d} \e^{\cdt \ddt} \e^{\r \s}
  \e^{\rdt \sdt} \e^{\m \n} \e^{\mdt \ndt} \notag \\
&+(((\e^{\adt \cdt} \e^{\ddt \bdt} \e^{\rdt \mdt} \e^{\ndt \sdt} \e^{\a
  \m} \e^{\b \n} \e^{\c \r} \e^{\d \s} + (\a \c , \b \d)) + (\c \cdt,
  \d \ddt))+(\m \mdt, \n \ndt))
\bigr)
\end{align}


\begin{thebibliography}{99}



%\cite{Maldacena:1997re}
\bibitem{Maldacena:1997re}
  J.~M.~Maldacena,
  %``The large N limit of superconformal field theories and supergravity,''
  Adv.\ Theor.\ Math.\ Phys.\  {\bf 2} (1998) 231
  [Int.\ J.\ Theor.\ Phys.\  {\bf 38} (1999) 1113]
  [arXiv:hep-th/9711200].
  %%CITATION = HEP-TH 9711200;%% 

%\cite{Witten:1998qj}
\bibitem{Witten:1998qj}
  E.~Witten,
  %``Anti-de Sitter space and holography,''
  Adv.\ Theor.\ Math.\ Phys.\  {\bf 2} (1998) 253
  [arXiv:hep-th/9802150].
  %%CITATION = HEP-TH 9802150;%%

%\cite{Gubser:1998bc}
\bibitem{Gubser:1998bc}
  S.~S.~Gubser, I.~R.~Klebanov and A.~M.~Polyakov,
  %``Gauge theory correlators from non-critical string theory,''
  Phys.\ Lett.\ B {\bf 428} (1998) 105
  [arXiv:hep-th/9802109].
  %%CITATION = HEP-TH 9802109;%%

%\cite{Eden:1998hh}
\bibitem{Eden:1998hh}
  B.~Eden, P.~S.~Howe, C.~Schubert, E.~Sokatchev and P.~C.~West,
% ``Four-point functions in N = 4 supersymmetric Yang-Mills theory at
%  two loops,''
  Nucl.\ Phys.\ B {\bf 557} (1999) 355
  [arXiv:hep-th/9811172].
  %%CITATION = HEP-TH 9811172;%%

%\cite{Eden:1999kh}
\bibitem{Eden:1999kh}
  B.~Eden, P.~S.~Howe, C.~Schubert, E.~Sokatchev and P.~C.~West,
%  ``Simplifications of four-point functions in N = 4 supersymmetric
%  Yang-Mills theory at two loops,''
  Phys.\ Lett.\ B {\bf 466} (1999) 20
  [arXiv:hep-th/9906051].
  %%CITATION = HEP-TH 9906051;%%

%\cite{Bianchi:1999ge}
\bibitem{Bianchi:1999ge}
  M.~Bianchi, S.~Kovacs, G.~Rossi and Y.~S.~Stanev,
  %``On the logarithmic behavior in N = 4 SYM theory,''
  JHEP {\bf 9908} (1999) 020
  [arXiv:hep-th/9906188].
  %%CITATION = HEP-TH 9906188;%%

%\cite{Eden:2000mv}
\bibitem{Eden:2000mv}
  B.~Eden, C.~Schubert and E.~Sokatchev,
%  ``Three-loop four-point correlator in N = 4 SYM,''
  Phys.\ Lett.\ B {\bf 482} (2000) 309
  [arXiv:hep-th/0003096].
  %%CITATION = HEP-TH 0003096;%%

%\cite{Bianchi:2000hn}
\bibitem{Bianchi:2000hn}
  M.~Bianchi, S.~Kovacs, G.~Rossi and Y.~S.~Stanev,
%  ``Anomalous dimensions in N = 4 SYM theory at order g**4,''
  Nucl.\ Phys.\ B {\bf 584} (2000) 216
  [arXiv:hep-th/0003203].
  %%CITATION = HEP-TH 0003203;%%

%\cite{Bianchi:2001cm}
\bibitem{Bianchi:2001cm}
  M.~Bianchi, S.~Kovacs, G.~Rossi and Y.~S.~Stanev,
  %``Properties of the Konishi multiplet in N = 4 SYM theory,''
  JHEP {\bf 0105} (2001) 042
  [arXiv:hep-th/0104016].
  %%CITATION = HEP-TH 0104016;%%

%\cite{Liu:1998ty}
\bibitem{Liu:1998ty}
  H.~Liu and A.~A.~Tseytlin,
% ``On four-point functions in the CFT/AdS correspondence,''
  Phys.\ Rev.\ D {\bf 59} (1999) 086002
  [arXiv:hep-th/9807097].
  %%CITATION = HEP-TH 9807097;%%

%\cite{Freedman:1998bj}
\bibitem{Freedman:1998bj}
  D.~Z.~Freedman, S.~D.~Mathur, A.~Matusis and L.~Rastelli,
  %``Comments on 4-point functions in the CFT/AdS correspondence,''
  Phys.\ Lett.\ B {\bf 452} (1999) 61
  [arXiv:hep-th/9808006].
  %%CITATION = HEP-TH 9808006;%%

%\cite{D'Hoker:1999jc}
\bibitem{D'Hoker:1999jc}
  E.~D'Hoker, D.~Z.~Freedman, S.~D.~Mathur, A.~Matusis and L.~Rastelli,
  %``Graviton and gauge boson propagators in AdS(d+1),''
  Nucl.\ Phys.\ B {\bf 562} (1999) 330
  [arXiv:hep-th/9902042].
  %%CITATION = HEP-TH 9902042;%%

%\cite{D'Hoker:1999pj}
\bibitem{D'Hoker:1999pj}
  E.~D'Hoker, D.~Z.~Freedman, S.~D.~Mathur, A.~Matusis and L.~Rastelli,
  %``Graviton exchange and complete 4-point functions in the AdS/CFT
  %  correspondence,'' 
  Nucl.\ Phys.\ B {\bf 562} (1999) 353
  [arXiv:hep-th/9903196].
  %%CITATION = HEP-TH 9903196;%%

%\cite{Arutyunov:1999fb}
\bibitem{Arutyunov:1999fb}
  G.~Arutyunov and S.~Frolov,
  %``Scalar quartic couplings in type IIB supergravity on AdS(5) x S(5),''
  Nucl.\ Phys.\ B {\bf 579} (2000) 117
  [arXiv:hep-th/9912210].
  %%CITATION = HEP-TH 9912210;%%

%\cite{Arutyunov:2000py}
\bibitem{Arutyunov:2000py}
  G.~Arutyunov and S.~Frolov,
  %``Four-point functions of lowest weight CPOs in N = 4 SYM(4) in
  %supergravity approximation,''
  Phys.\ Rev.\ D {\bf 62} (2000) 064016
  [arXiv:hep-th/0002170].
  %%CITATION = HEP-TH 0002170;%%
  
%\cite{Arutyunov:2000ku}
\bibitem{Arutyunov:2000ku}
  G.~Arutyunov, S.~Frolov and A.~C.~Petkou,
%  ``Operator product expansion of the lowest weight CPOs in N = 4  SYM(4) at
%  strong coupling,''
  Nucl.\ Phys.\ B {\bf 586} (2000) 547
  [Erratum-ibid.\ B {\bf 609} (2001) 539]
  [arXiv:hep-th/0005182].
  %%CITATION = HEP-TH 0005182;%%

%\cite{Arutyunov:2001mh}
\bibitem{Arutyunov:2001mh}
  G.~Arutyunov, B.~Eden, A.~C.~Petkou and E.~Sokatchev,
  % ``Exceptional non-renormalization properties and OPE analysis of chiral
  %four-point functions in N = 4 SYM(4),''
  Nucl.\ Phys.\ B {\bf 620} (2002) 380
  [arXiv:hep-th/0103230].
  %%CITATION = HEP-TH 0103230;%%

%\cite{Dolan:2001tt}
\bibitem{Dolan:2001tt}
  F.~A.~Dolan and H.~Osborn,
  % ``Superconformal symmetry, correlation functions and the operator product
  %expansion,''
  Nucl.\ Phys.\ B {\bf 629} (2002) 3
  [arXiv:hep-th/0112251].
  %%CITATION = HEP-TH 0112251;%%

%\cite{Dolan:2004iy}
\bibitem{Dolan:2004iy}
  F.~A.~Dolan and H.~Osborn,
  %``Conformal partial wave expansions for N = 4 chiral four point
%  functions,'' 
  Annals Phys.\  {\bf 321} (2006) 581
  [arXiv:hep-th/0412335].
  %%CITATION = HEP-TH 0412335;%%

%\cite{Eden:2000bk}
\bibitem{Eden:2000bk}
  B.~Eden, A.~C.~Petkou, C.~Schubert and E.~Sokatchev,
%   ``Partial non-renormalisation of the stress-tensor four-point
%  function in  N = 4 SYM and AdS/CFT,''
  Nucl.\ Phys.\ B {\bf 607} (2001) 191
  [arXiv:hep-th/0009106].
  %%CITATION = HEP-TH 0009106;%%

%\cite{Dolan:2000ut}
\bibitem{Dolan:2000ut}
  F.~A.~Dolan and H.~Osborn,
  %``Conformal four point functions and the operator product expansion,''
  Nucl.\ Phys.\ B {\bf 599} (2001) 459
  [arXiv:hep-th/0011040].
  %%CITATION = HEP-TH 0011040;%%

%\cite{Heslop:2002hp}
\bibitem{Heslop:2002hp}
  P.~J.~Heslop and P.~S.~Howe,
  %``Four-point functions in N = 4 SYM,''
  JHEP {\bf 0301} (2003) 043
  [arXiv:hep-th/0211252].
  %%CITATION = HEP-TH 0211252;%%

%\cite{Davydychev:1992mt}
\bibitem{Davydychev:1992mt}
  A.~I.~Davydychev and J.~B.~Tausk,
  % ``Two loop selfenergy diagrams with different masses and the momentum
  %expansion,''
  Nucl.\ Phys.\ B {\bf 397} (1993) 123.
  %%CITATION = NUPHA,B397,123;%%

%\cite{Dolan:2004mu}
\bibitem{Dolan:2004mu}
  F.~A.~Dolan, L.~Gallot and E.~Sokatchev,
  %``On four-point functions of 1/2-BPS operators in general dimensions,''
  JHEP {\bf 0409} (2004) 056
  [arXiv:hep-th/0405180].
  %%CITATION = HEP-TH 0405180;%%

%\cite{Usyukina:1992jd}
\bibitem{Usyukina:1992jd}
  N.~I.~Usyukina and A.~I.~Davydychev,
  %``An Approach to the evaluation of three and four point ladder diagrams,''
  Phys.\ Lett.\ B {\bf 298} (1993) 363.
  %%CITATION = PHLTA,B298,363;%%


\end{thebibliography}
\end{document}